\documentclass[letterpaper, 10pt, conference]{ieeeconf}      

\IEEEoverridecommandlockouts                              

\usepackage{amsmath,nccmath,graphicx,epstopdf}
\usepackage[space]{cite}
\usepackage[binary-units=true,per-mode=symbol]{siunitx}
\usepackage[colorinlistoftodos]{todonotes}
\usepackage{subfigure, multirow, soul, hyperref}

\definecolor{ascolor}{rgb}{.1, 0.7, 0.1}

\title{\LARGE \bf
Adaptive Video Configuration and Bitrate Allocation\\for Teleoperated Vehicles
}

\author{Andreas Schimpe, Simon Hoffmann and Frank Diermeyer%
\thanks{The authors are with the Institute for Automotive Technology at the Technical University of Munich (TUM), 85748 Garching bei M\"unchen, Germany. {\tt\small andreas.schimpe@tum.de}}%
}

\usepackage{textcomp}
\usepackage{lipsum}
\newcommand\copyrighttext{%
\footnotesize \textcopyright 2021 IEEE. Personal use of this material is permitted. Permission from IEEE must be obtained for all other uses, in any current or future media, including reprinting/republishing this material for advertising or promotional purposes, creating new collective works, for resale or redistribution to servers or lists, or reuse of any copyrighted component of this work in other works.
}
\newcommand\copyrightnotice{%
    \begin{tikzpicture}[remember picture,overlay]
    \node[anchor=south,yshift=10pt, xshift=10pt] at (current page.south) {\fbox{\parbox{\dimexpr\textwidth-\fboxsep-\fboxrule\relax}{\copyrighttext}}};
    \end{tikzpicture}%
}

\begin{document}
\maketitle
\copyrightnotice 
\thispagestyle{empty}
\pagestyle{empty}

\begin{abstract}
Vehicles with autonomous driving capabilities are present on public streets. However, edge cases remain that still require a human in-vehicle driver. Assuming the vehicle manages to come to a safe state in an automated fashion, teleoperated driving technology enables a human to resolve the situation remotely by a control interface connected via a mobile network. While this is a promising solution, it also introduces technical challenges, one of them being the necessity to transmit video data of multiple cameras from the vehicle to the human operator. In this paper, an adaptive video streaming framework specifically designed for teleoperated vehicles is proposed and demonstrated. The framework enables automatic reconfiguration of the video streams of the multi-camera system at runtime. Predictions of variable transmission service quality are taken into account. With the objective to improve visual quality, the framework uses so-called rate-quality models to dynamically allocate bitrates and select resolution scaling factors. Results from deploying the proposed framework on an actual teleoperated driving system are presented. 
\end{abstract}


\section{Introduction}
Great effort has been put towards the development of fully automated and driverless vehicles. In particular, improvements have been made in perception systems, planning, and control algorithms. Currently, automated vehicles are already being tested with safety drivers on public streets. Assuming the automation detects that a traffic situation cannot be resolved independently, teleoperated driving offers a solution.

\subsection{Teleoperated Driving}
With teleoperated driving~(ToD) technology, a vehicle can be controlled remotely. Sensor and vehicle data, e.g., video feeds, are transmitted via mobile networks from the vehicle to a remote control center. There, the data are displayed to a human operator, who generates control commands. These are then transmitted back to the vehicle for execution. \\
ToD comes with a number of challenges. First of all, the teleoperation is subject to latency due to delays in system components and transmission of the data via a mobile network. However, with advances in computational power, sensor and actuator technologies, delays can be reduced significantly~\cite{Georg2020a}. In addition, novel mobile network standards promise even greater reduction~\cite{Kousaridas2019}. Situational awareness of the operator poses another great challenge to ToD. Not being physically located in the vehicle, the perception of the operator is primarily based on multiple video streams of the cameras on the vehicle. A display based on a spherical video canvas is proposed in~\cite{Georg2019}. It is applied in a long-term study~\cite{Georg2020b}, reporting an increase in the operator immersion. The study also indicates an influence of different video quality settings. With the objective to maximize the visual quality of the video streams, this paper addresses the task of automated dynamic adaptation of video stream parameters, in dependence on predictions of bitrate availability.

\subsection{Predictive Quality of Service}
\label{ch:pQoS}
Through predictive Quality of Service~(pQoS), an application can receive predictions of the communication quality of a mobile network, e.g., available bitrate. It enables proactive adaptation of the data transmission, instead of reacting to occurrences of increasing network latencies or packet loss. Predicting communication quality is affected by multiple parameters~\cite{Akselrod2017}. If~pQoS is not provided by the mobile network operator, data- and machine learning-based models can be used to generate predictions~\cite{Jomrich2018}. Coupled with 5G, the topic attracts a lot of interest in industry, especially for automotive applications~\cite{Kousaridas2019, 5GAApQoS2020}. With demand for low latency and high data rates on the mobile network link, ToD is an application that can greatly benefit of pQoS. The framework in this paper assumes that predictions of the allocatable uplink bitrate are available and will be used to perform video stream adaptations of the ToD system.

\subsection{Adaptive Multi-Camera Video Streaming}
While video streaming is well-established in many domains, teleoperation, and therein especially ToD, is a video streaming application with special demands. As the operator controls the vehicle remotely in real time, the video streams must be delivered with the lowest possible delay. Therefore, it is not possible to make use of the full compression capabilities of nowadays codecs. Also, the bandwidth of the mobile network is limited and varies. These factors compromise the achievable visual quality and make adaptive video streaming necessary. Over the years, adaptive video streaming was primarily investigated for the purpose of entertainment services. For instance, the approach in~\cite{Chen2017} optimizes playback bitrates using pre-generated rate-quality models and other statistics. Other rate control methods, such as look-ahead approaches~\cite{Perkins1995, Wang2010} enable variable allocation of a constant total bitrate for multiple videos. \\
In the field of teleoperation, the work presented in~\cite{Johansson2018} performs variable bitrate allocation for a multi-camera system at varying radio conditions. This is related to some of the objectives in this paper. However, the camera weighting was based more on an intuition for the priorities of the camera. The resulting visual quality of the videos was not taken into account. A thorough assessment of visual quality in the ToD simulation setup TELECARLA~\cite{Hofbauer2020a} is presented in~\cite{Hofbauer2020b}. Therein, the video streams are parametrized in dependence of their orientation and the driving scenario. Dynamic adjustment of the video parameters is not considered.

\subsection{Contributions}
In this paper, a video streaming framework is proposed that aims at providing the flexibility as required by ToD. This includes variable prioritization of cameras and dynamic adjustment of video parameters. The videos can be parametrized in different ways. The framework either takes in manual input from the operator or predictions of the currently available bitrate. In the latter case, bitrates are allocated and resolution scaling factors are selected automatically. This is performed based on so-called rate-quality models, which were generated for each camera of the system. Insights into the deployment of the proposed framework on an actual ToD system with eight cameras are given. The generated rate-quality models are discussed and the functionality of the framework during operation is demonstrated in an experimental test drive. The source code of the framework is open-source and available as a Robot Operating System~(ROS) package on GitHub\footnote{\href{https://github.com/TUMFTM/tod\_perception}{https://github.com/TUMFTM/tod\_perception}}.

\section{Adaptive Video Streaming Framework}
\begin{figure}[!t]
    \includegraphics[width=\linewidth]{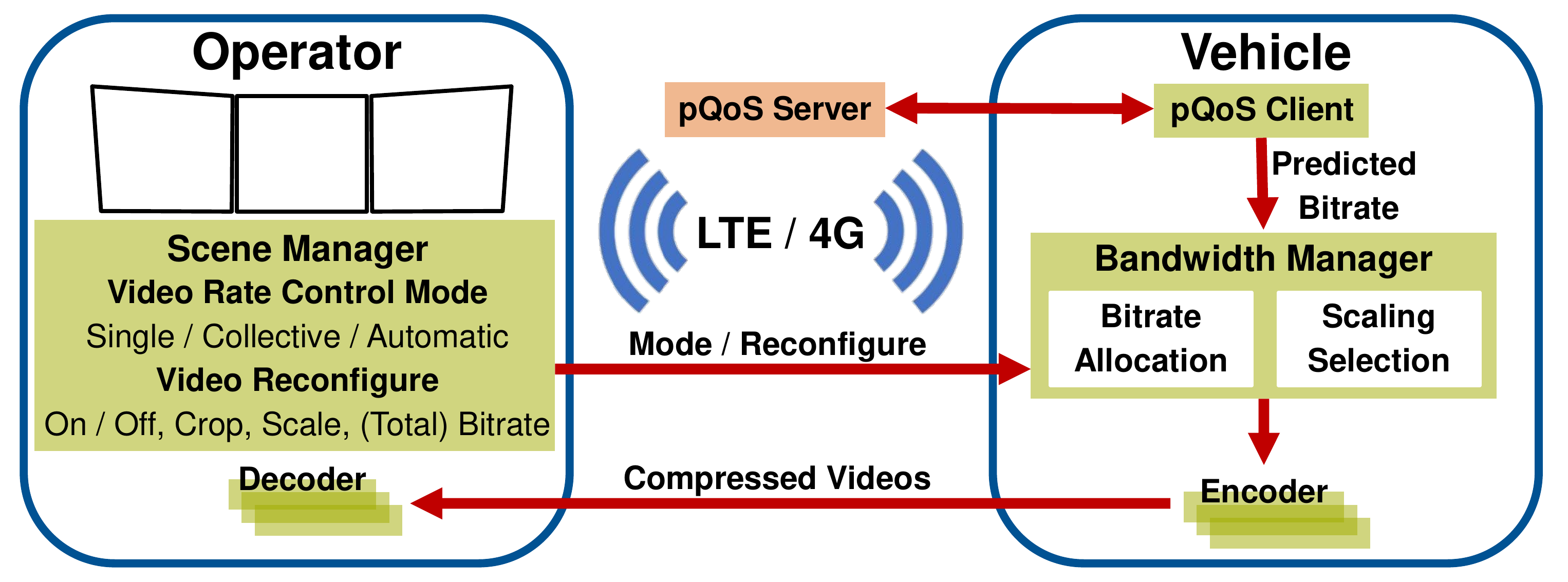}
    \caption{Architecture of Adaptive Video Streaming Framework}
    \label{fig:vidArch}
\end{figure}
To control and adapt multiple video streams, the architecture, shown in Fig.~\ref{fig:vidArch}, is proposed. For each camera on the vehicle, the raw pixel data of the videos are compressed by an encoder. The compressed data are transmitted via a mobile network to the operator side, where they are decoded and displayed. A scene manager user interface offers options for the operator to switch between three video rate control modes: Single~(S), Automatic~(A) and Collective~(C). In mode~S, the videos can be turned on and off, or reconfigured individually. This means that the region of interest~(ROI), characterized by width, height, horizontal and vertical offset in pixel, can be set for each video. Furthermore, the video resolution scaling factor and the target bitrate of the encoder can be set. In mode A, the ROIs are kept constant, i.e., as the operator has set them in mode~S. Based on predictions of the (total) bitrate available from the pQoS client, the bitrate allocation and resolution scaling factor selection for the individual video streams are performed and applied by the bandwidth manager automatically. Video rate control mode~C is similar to mode~A, with the difference that the total bitrate is not given by the pQoS client, but by the operator through the scene manager. In the following, the procedures of bitrate allocation and resolution scaling factor selection, as performed by the bandwidth manager, are described in detail.

\subsection{Bitrate Allocation}
Given the operator's selected regions of interest and a prediction of the current available bitrate, each camera of the multi-view video system is allocated a bitrate. At first, from the ROI for each camera~$i$, the bitrate demand~$b_{\textrm{dem},i}$ is computed by
\begin{equation}
b_{\textrm{dem},i} = \frac{p_{\textrm{set},i}}{p_{\textrm{full},i}} \, b_{\textrm{full},i} , 
\label{eq:optBr}
\end{equation}
where~$p_{(\cdot)} = w_{(\cdot)} \, h_{(\cdot)}$, with image width~$w_{(\cdot)}$ and height~$h_{(\cdot)}$, are the amount of pixels of the ROI, set by the operator, and the full image of the camera, respectively. For a camera that is turned off, $p_{\textrm{set},i}$ is set to zero. The maximum bitrate that should be allocated to the camera for the full image is denoted by~$b_{\textrm{full},i}$. This parameter, implicitly formulating a weight of the respective camera, is set manually and will be discussed in more detail in Sec.~\ref{ch:resQualMdls}. In addition, the bitrate allocation can be complemented by a set of explicit camera weights. \\
From the bitrate demand, the allocated bitrate~$b_{\textrm{alloc},i}$ is calculated as
\begin{equation}
b_{\textrm{alloc},i} = \frac{b_{\textrm{dem},i}}{\sum_{j=1}^{N} b_{\textrm{dem},j}} \, b_{\textrm{pred}},
\label{eq:allocBr}
\end{equation}
whereby~$N$ and~$b_{\textrm{pred}}$ denote the total number of cameras and the prediction of the bitrate available from the pQoS client, respectively.

\subsection{Resolution Scaling Selection}
From the set ROI and allocated bitrate for each camera, the resolution scaling factor that maximizes the average visual quality is selected for each camera. This selection is based on rate-quality models. Their generation and metrics for the visual quality are described in more detail in the next section. For better readability, the camera index~$i$ is omitted in this section. \\
Generally speaking, for each camera, the rate-quality model yields a set~$S$ of resolution scaling factors
\begin{equation}
S = \{ s_{1}, \; s_{2} \; ... \; s_{R} \}. 
\end{equation}
Within~$S$, the resolution scaling factors~$s_r$ are~$R$ real numbers in the range~$] \, 0, \, 1 \, ]$. Each factor maximizes the average visual quality in a range~$B_r$ of encoder target bitrates. \\
Given the allocated bitrate for a camera from~\eqref{eq:allocBr}, the procedure selects the factor~$s_r$ for which
\begin{equation}
b_{\text{alloc}} \frac{p_{\text{full}}}{p_{\text{set}}} \in B_r = \begin{cases}
	[ \, b_{\text{min,r}}, \; b_{\text{min},r+1}  \, [ & , \text{if} \; r < R \\
	[ \, b_{\text{min},R}, \; b_{\text{full}} \, ] & , \text{if} \; r = R
\end{cases}
\end{equation}
holds. This aims at incorporating the selection of a greater resolution scaling factor in case the operator has set an ROI smaller than the full image of the camera.

\section{Generation of Rate-Quality Models}
\label{ch:modelGen}

\begin{figure}[!t]
    \includegraphics[width=\linewidth]{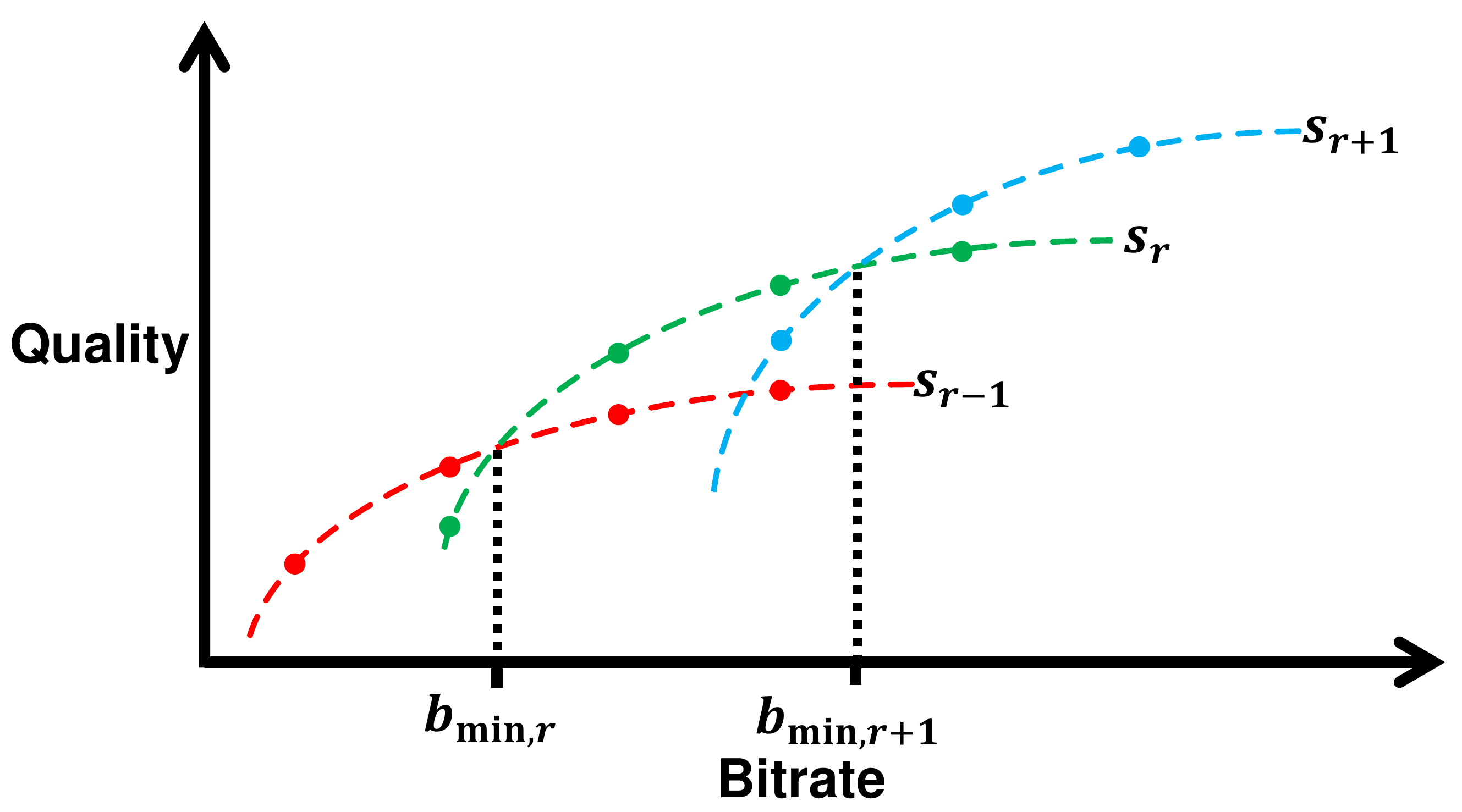}
    \caption{Conceptual Illustration of the Rate-Quality Model}
    \label{fig:descrRateQualCurves}
\end{figure}%

To perform the previously described resolution scaling factor selection, rate-quality models are generated beforehand. This follows a concept presented in~\cite{Chen2017}, which splits a video in five second, non-overlapping chunks. Each chunk is encoded at different resolutions, and compression rates. The average visual quality of each chunk is stored in a database. These data then enable the selection of parameters that maximize the visual quality of each chunk during playback. Fig.~\ref{fig:descrRateQualCurves} conceptually illustrates the rate-quality model for three resolutions, i.e., scaling factors~$s_r$. Each curve represents the visual quality of the video at a certain scaling factor over varying bitrate. Each scaling factor achieves the highest visual quality in a certain bitrate range. From the points of intersection of the rate-quality curves, the values~$b_{\text{min},r}$ are determined. \\
In ToD, the streamed videos are not recordings, but live feeds. In consequence, the described concept is adapted to the problem at hand. For the presented adaptive video streaming framework, rate-quality models are not obtained for short video chunks, but for each camera of the ToD system. Therefor, a representative driving sequence with straight sections and turns is recorded with each camera in a raw, non-compressed format. This is then downscaled and encoded with a constant encoder target bitrate. Afterwards, the compressed video is decoded and scaled up to its original resolution. From this, the visual quality of the video frame is assessed through a full-reference metric. Well known metrics are the Peak Signal-to-Noise Ratio~(PSNR) or the Mean Structural Similarity~(MSSIM)~\cite{Wang2004}. Alternatively, a no-reference image quality assessment method, such as the Blind/Referenceless Image Spatial Quality Evaluator~(BRISQUE)~\cite{Mittal2012}, may be used. Recently, also Netflix' Video Multi-Method Assessment Fusion~(VMAF)~\cite{VMAF2018} has gained in popularity. The visual quality is computed for each video frame of the driving sequence. Finally, the average over all frames represents one data point of the rate-quality model. The procedure is repeated for multiple resolution scaling factors and encoder target bitrates. This yields the full camera rate-quality model, which is used in the adaptive video streaming framework, as described in the previous section.

\section{Experimental System Setup}

\begin{figure}[!t]
    \includegraphics[width=\linewidth]{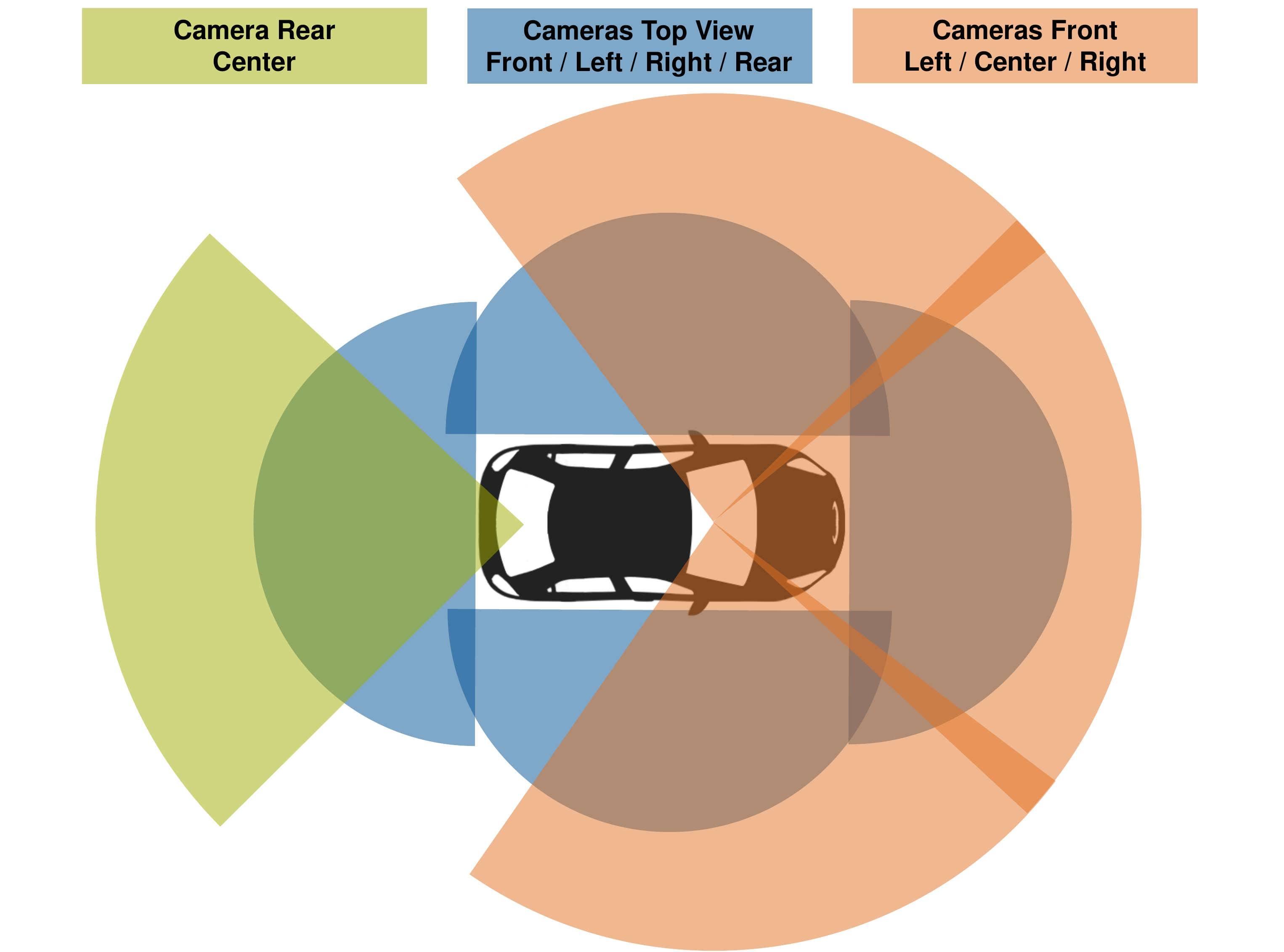}
    \caption{Camera Setup and Fields of View on Experimental Vehicle}
    \label{fig:camViews}
\end{figure}%

The video streaming framework presented in this paper has been deployed on the experimental vehicle for ToD described in~\cite{Georg2019}. Since then, the camera setup has been complemented by two front-mounted cameras. As illustrated in Fig.~\ref{fig:camViews}, the setup now consists of a total of eight cameras. These are
\begin{itemize}
\item three front-mounted and one rear-mounted camera operating at 40 Hz, 
\item and four top view cameras operating at 30 Hz, with an opening angle of~$\SI{180}{\deg}$, each for monitoring the close surroundings of the vehicle. 
\end{itemize}
The vehicle camera parameters are summarized in~Tab.~\ref{tab:camParams}. The reported resolution values partially do not correspond to the native camera resolutions, but already incorporate cropping of the videos due to overlap or regions, such as the sky, which certainly do not contain valuable information, and thus do not need to be transmitted to the operator. The obtained values of rate-quality models are discussed in more detail in the next section. \\
The adaptive video streaming framework is implemented within ROS~\cite{Quigley2009}. The video compression and transmission is handled using GStreamer~\cite{GStreamer}. The GStreamer Real-Time Streaming Protocol~(RTSP) Server Library~\cite{RtspServer}, implementing an RTSP-based client-server model, is used for establishing and controlling the video streaming sessions between the operator and the vehicle. GStreamer is a modular framework based on so-called plugins, each providing a certain functionality. By putting plugins together in a pipeline, different multimedia streaming applications can be created. \\
Within the presented video streaming framework, the pipeline for each camera on the vehicle side consists of the following plugins:
\begin{center}
\texttt{appsrc} $\rightarrow$ \texttt{videocrop} \\
$\rightarrow$ \texttt{videoscale} $\rightarrow$ \texttt{capsfilter} $\rightarrow$ \texttt{x264enc} \\
$\rightarrow$ \texttt{rtph264pay} $\rightarrow$ \texttt{rtsp server}. 
\end{center} 
From the ROS image callback, the raw video frames are pushed to the \texttt{appsrc}. The \texttt{x264enc} compresses the raw video using the H.264 codec~\cite{H264}. For the transmission via the \texttt{rtsp server}, the compressed data is split into Real-Time Transport Protocol~(RTP) packets by the \texttt{rtph264pay}. Adapting the configuration and control of the video stream at runtime happens through parametrization of certain plugins. These are the \texttt{videocrop} element to adapt the ROI, the \texttt{capsfilter} element to force the \texttt{videoscale} to perform a certain resolution scaling, and the \texttt{x264enc} to encode the video stream at the (target) bitrate. \\
On the operator side, each video is received with the following pipeline:
\begin{center}
\texttt{rtspsrc} $\rightarrow$ \texttt{rtph264depay} $\rightarrow$ \texttt{avdec\_h264} \\
$\rightarrow$ \texttt{appsink},
\end{center}
where the \texttt{rtspsrc} creates the RTSP client that connects to the server to receive the RTP packets. These are passed on to the \texttt{rtph264depay} to reassemble the H.264 compressed data. The \texttt{avdec\_h264} decodes these into raw video frames, which are retrieved from the \texttt{appsink} and published to ROS for the operator display. \\
Several hardware and software design choices for the ToD system were driven by the goal to minimize the age of the information that is being displayed to the operator. An extensive and thorough analysis of hardware components and their latencies is presented in~\cite{Georg2020a}. The design choices include high camera frame rates and, despite lower compression rates, the use of the faster H.264 over H.265~\cite{H265}. With~H.264, a low latency is achieved by parametrizing the encoder to use the~\texttt{speed-preset=superfast} and the \texttt{tune=zero-latency} property. Also, to take advantage of the continuous, cut-free videos during ToD, the \texttt{intra-refresh} property is used to periodically distribute the refresh of the video key frames. \\
With the described parameter settings and other hardware component choices, such as gaming monitors with minimal input lag and high refresh rates, the latency of the video streaming system can be reduced effectively. With a wired connection, the end-to-end latency, measured from an event happening in front of the camera until it is being displayed on the operator monitor, is below \SI{120}{\milli\second} for the three front-mounted cameras~\cite{Georg2020a}. Although this paper does not explicitly aim at optimizing latency, these characteristics of the given ToD system are described here, as they significantly influence the obtained rate-quality models that are presented in the following section. 

\begin{table*}[!t]
\caption{Camera and Rate-Quality Model Parameters of Experimental Vehicle}
\centering
\begin{tabular}{| c | c | c | c | c | c | c |}
\hline
& Front Left / Right & Front Center & Rear Center & Top View Front & Top View Left / Right & Top View Rear \\\hline
$p_{\text{full}}$ (MPx) & \multicolumn{2}{c|}{$2.0$} & $2.3$ & \multicolumn{3}{c|}{$1.02$} \\\hline
$w_{\text{full}}$ (Px), $h_{\text{full}}$ (Px) & \multicolumn{2}{c|}{1920, 1040} & 1920, 1200 & \multicolumn{3}{c|}{1280, 800}\\\hline
$S$ & \multicolumn{3}{c|}{$\{ 0.125, 0.25, 0.5 \}$} & \multicolumn{3}{c|}{$\{ 0.25, 0.375, 0.5 \} $} \\\hline
$b_{\text{full}}$ (Mbit/s) & $6.0$ & $5.0$ & $4.0$ & $3.0$ & $3.0$ & $3.0$ \\\hline
$b_{\text{min},r}$ (Mbit/s) & $\{ 0, 0.3, 0.95 \}$ & $\{ 0, 0.2, 0.45 \}$ & $\{ 0, 0.4, 1.8 \}$ & $\{ 0, 0.15, 0.4 \}$ & $\{ 0, 0.15, 0.4 \}$ & $\{ 0, 0.2, 0.4 \}$ \\\hline
\end{tabular}
\label{tab:camParams}
\end{table*}%

\section{Results}
In this section, rate-quality models generated for the experimental vehicle are presented. Furthermore, results of an experimental drive give insights into the operation of the framework at runtime.

\subsection{Rate-Quality Models}
\label{ch:resQualMdls}

\begin{figure}[!t]
    \subfigure[Front Center] {
        \includegraphics[width=0.465\linewidth]{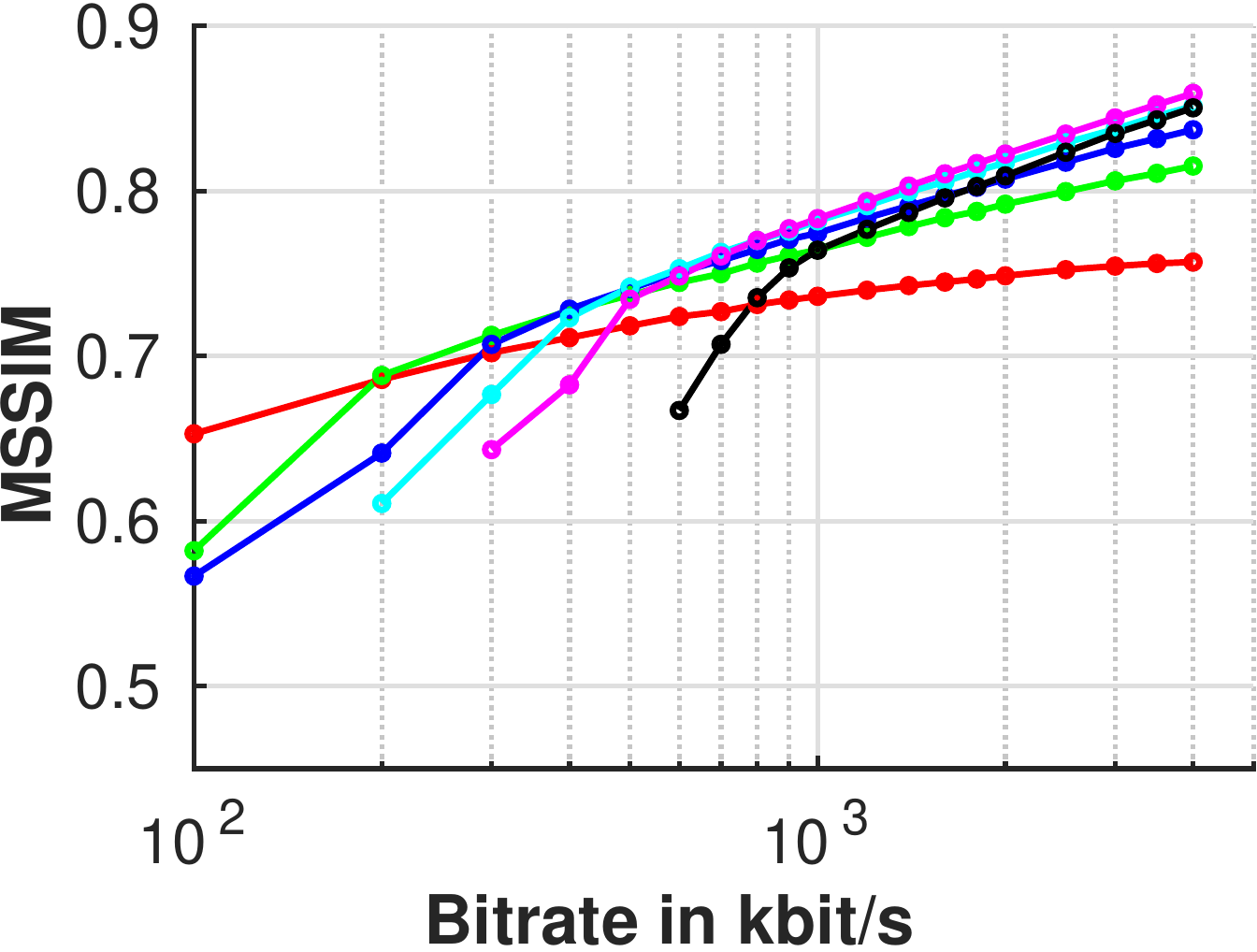}
    }
    \subfigure[Front Left and Right Average] {
        \includegraphics[width=0.465\linewidth]{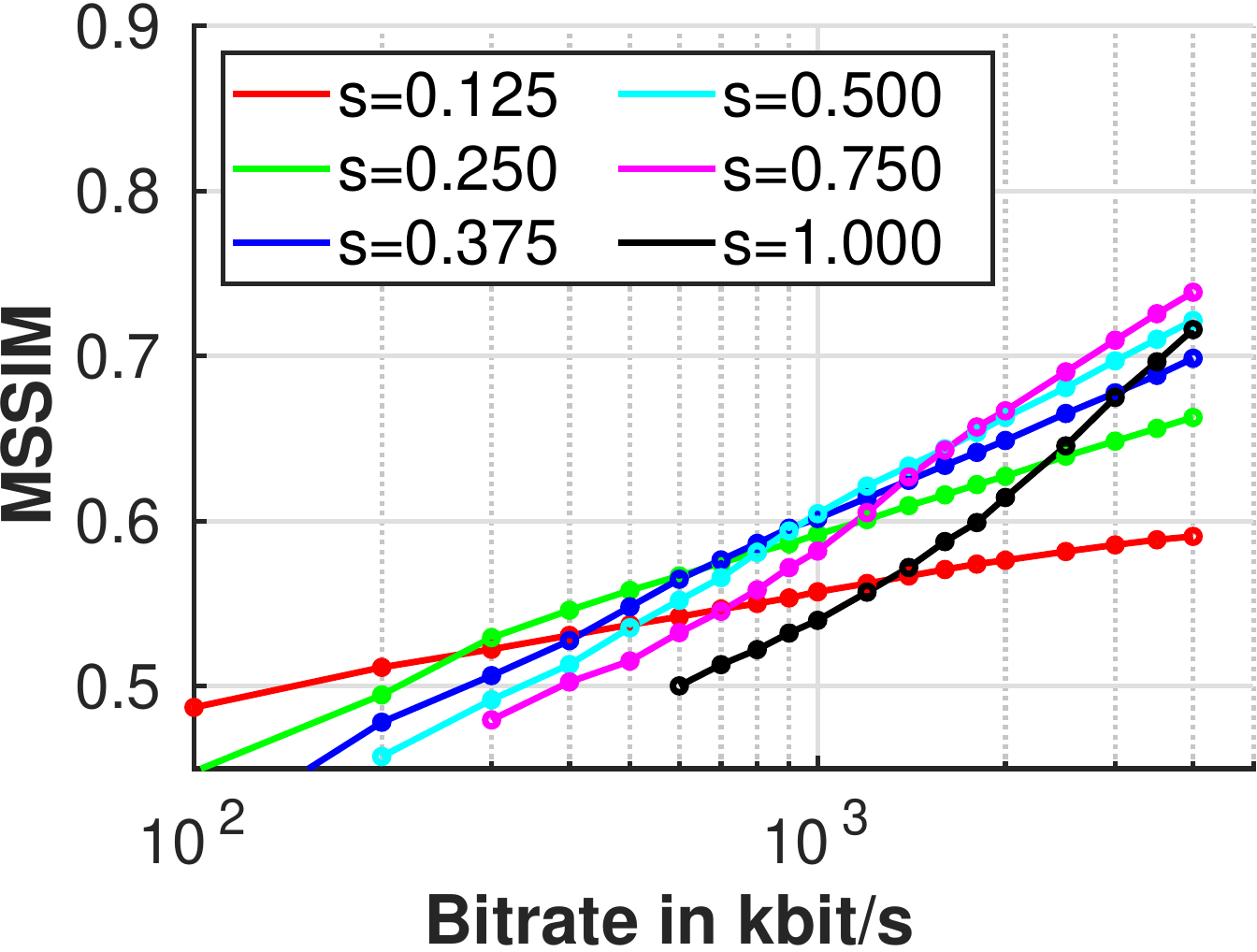}
    }
    \caption{Rate-Quality Models for Front-Mounted Cameras}
    \label{fig:qualCurves}
\end{figure}%

For the described camera setup, rate-quality models were generated with the methodology described in Sec.~\ref{ch:modelGen}. Sequences of approximately $40$~seconds during regular driving at moderate speeds were recorded in raw image format. The recordings were then run through the described compression-decompression pipeline at eight resolution scaling factors $S = \{ 0.125, 0.25, 0.375, ... \, 1.0 \}$ and encoder target bitrates between~$50$ and~\SI{4000}{\kilo\bit\per\second}. From the average MSSIMs for all parameter settings, resolution scaling factors that yield the highest quality in a certain bitrate range are selected. \\
The rate-quality models of the front-mounted center camera and the average of the front-mounted left and right cameras are shown in Fig.~\ref{fig:qualCurves}. All three cameras are of the same type. For better clarity, selected resolution scaling factors are displayed, only. It can be observed that each factor yields a bitrate range in which it maximizes the MSSIM, motivating dynamic scaling in the first place. Another observation is the major impact of the camera orientations on the quality model. For instance, as the motion in the videos of the side-facing cameras is greater, the bitrate ranges of the respective resolution scaling factors are different from the center-facing camera. Furthermore, overall lower quality scores are achieved. In consequence, this should be compensated for in the rate-quality model by assigning greater~$b_{\text{full}}$ to the side-facing cameras. Plots of the quality models for the other cameras can be found in Fig.~\ref{fig:otherQualCurves} in the Appendix. \\
From the rate-quality models, it is concluded that not all scaling factors are worthwhile to be applied. For instance, it turned out that some scaling factors are the best choice in a comparably narrow bitrate range, only. Also, resolution scaling factors close to~$1.0$, requiring greater computational power and encoding time, are only reasonable if the related camera bitrates are maintainable in the 4G/LTE network that is shared with multiple users. \\
The rate-quality model parameters used for the experimental vehicle are reported in Tab. I. Different resolution scaling factors turned out to be preferred for the different types of cameras. Also, scaling factors up to~$0.5$ came into use, only. However, with further advancements of mobile network standards and computing technologies, this is expected to change.

\subsection{Driving Test}
\begin{figure}[!t]
    \includegraphics[width=\linewidth]{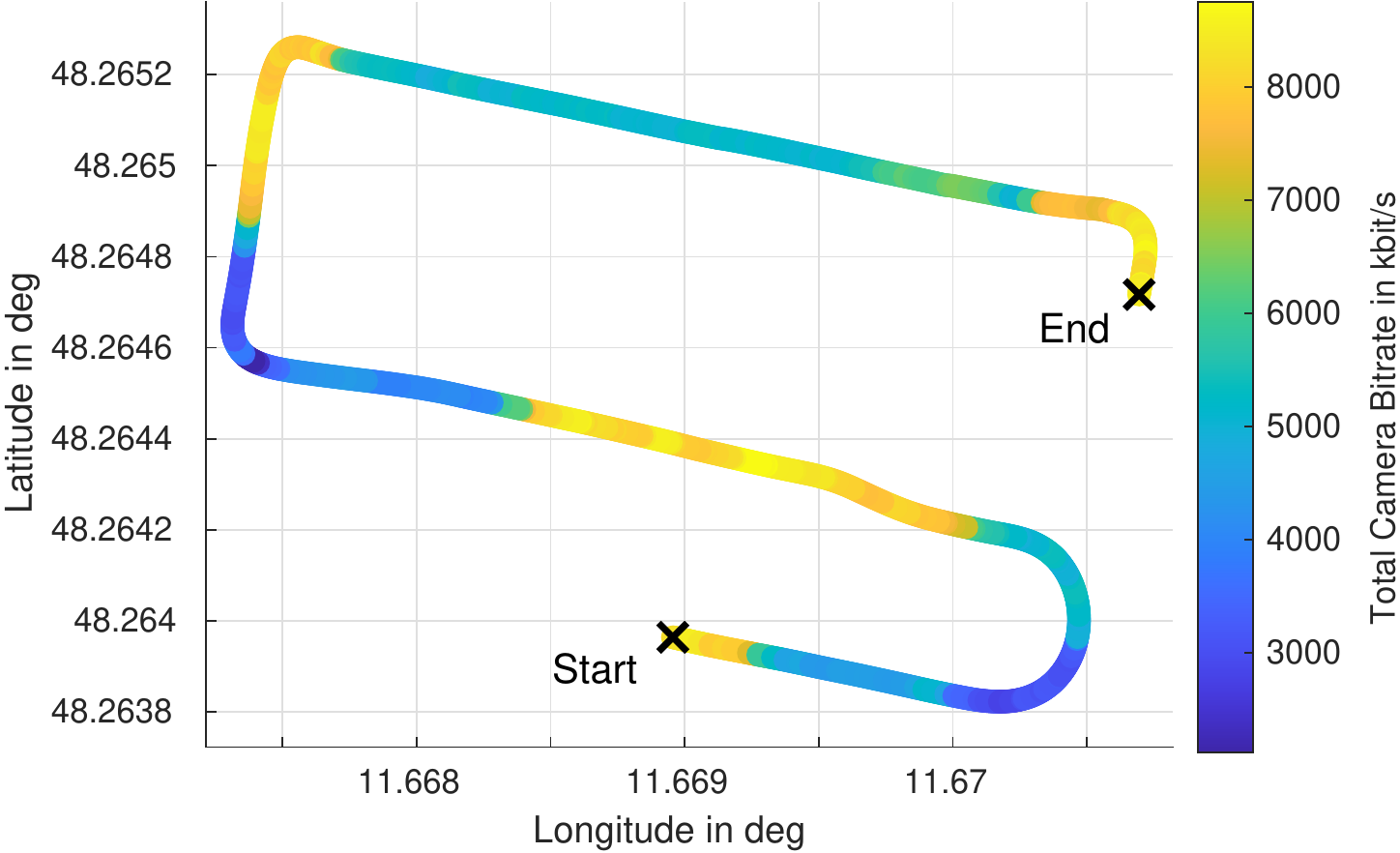}
    \caption{Actual Total Camera Bitrates over GPS Location of Vehicle}
    \label{fig:bitrateOnGps}
\end{figure}
\begin{figure}[!t]
    \subfigure[Front- and Rear-Mounted] {
        \includegraphics[width=0.465\linewidth]{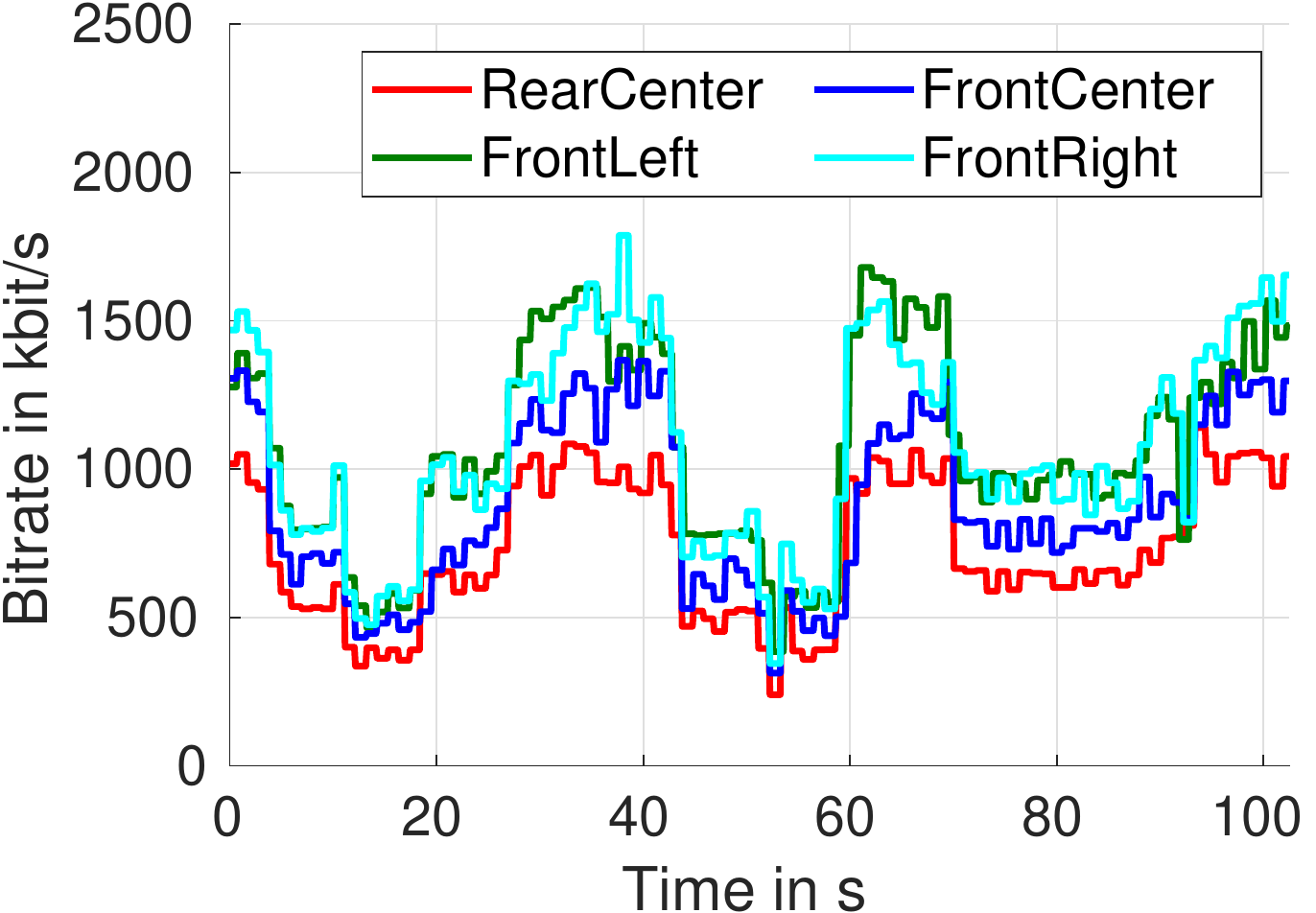}
    }
    \subfigure[Top Views] {
        \includegraphics[width=0.465\linewidth]{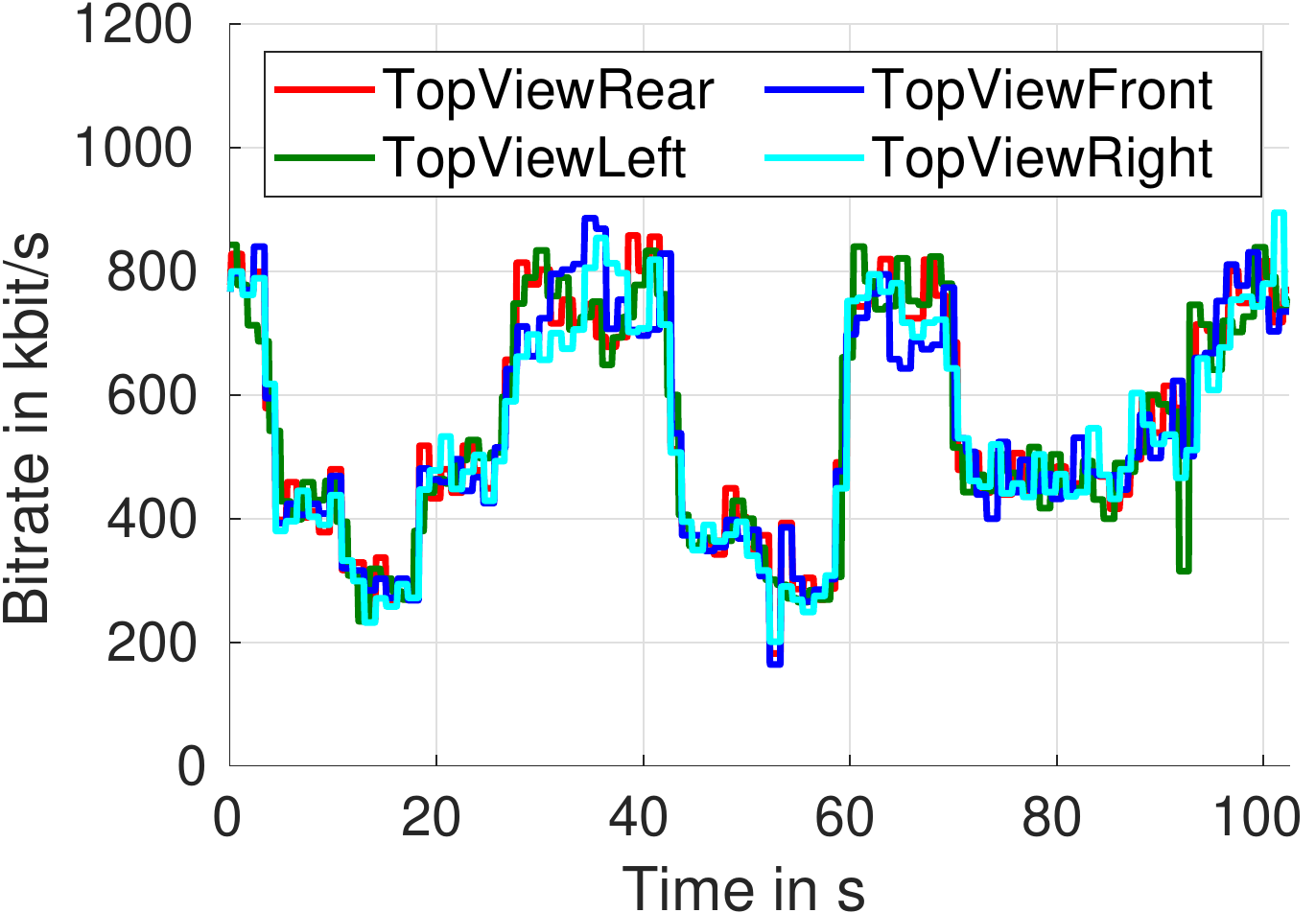}
    }
    \caption{Actual Camera Bitrates over Time}
    \label{fig:bitrateOverTime}
\end{figure}

\begin{figure}[!t]
    \subfigure[Front- and Rear Mounted] {
        \includegraphics[width=0.465\linewidth]{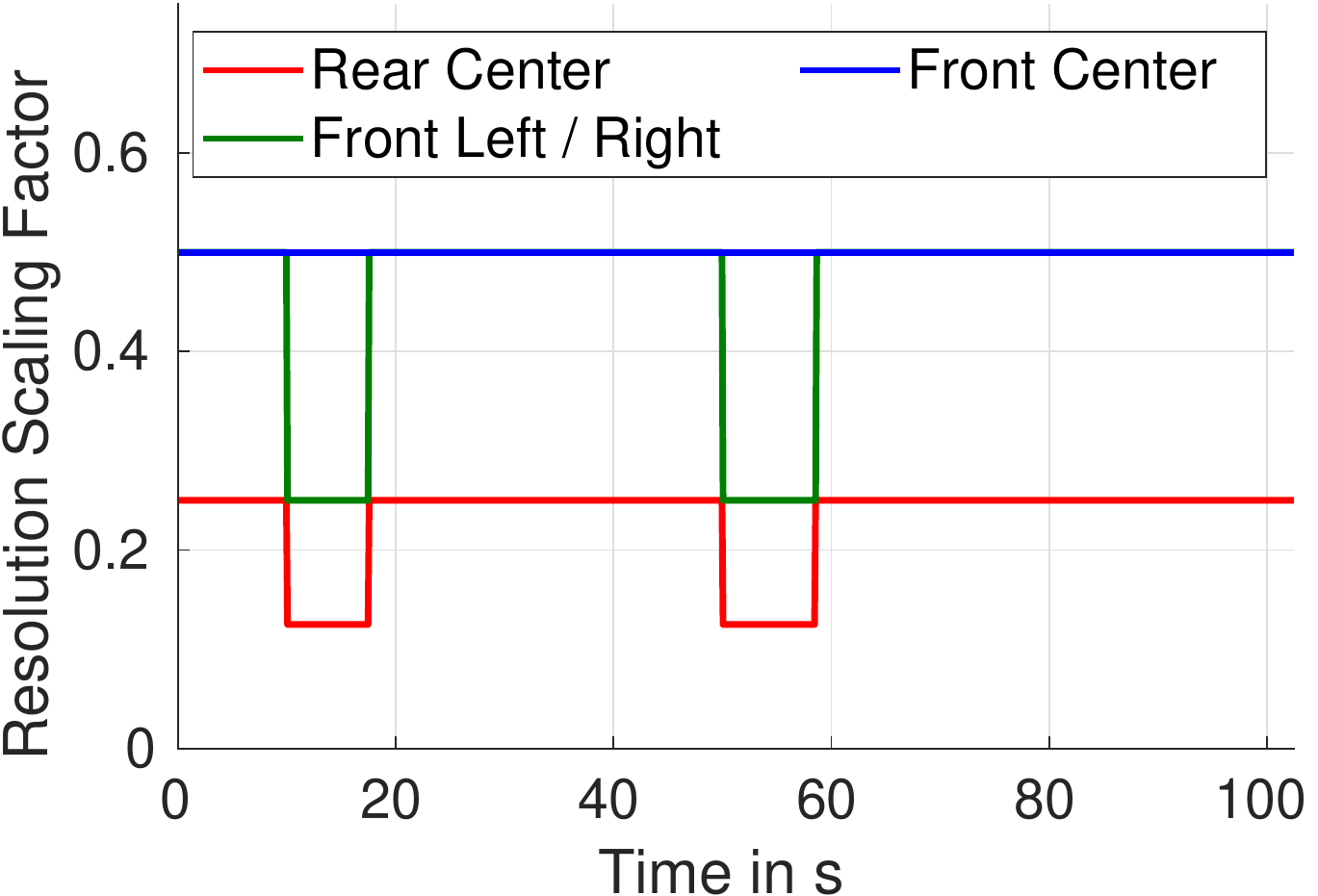}
    }
    \subfigure[Top Views] {
        \includegraphics[width=0.465\linewidth]{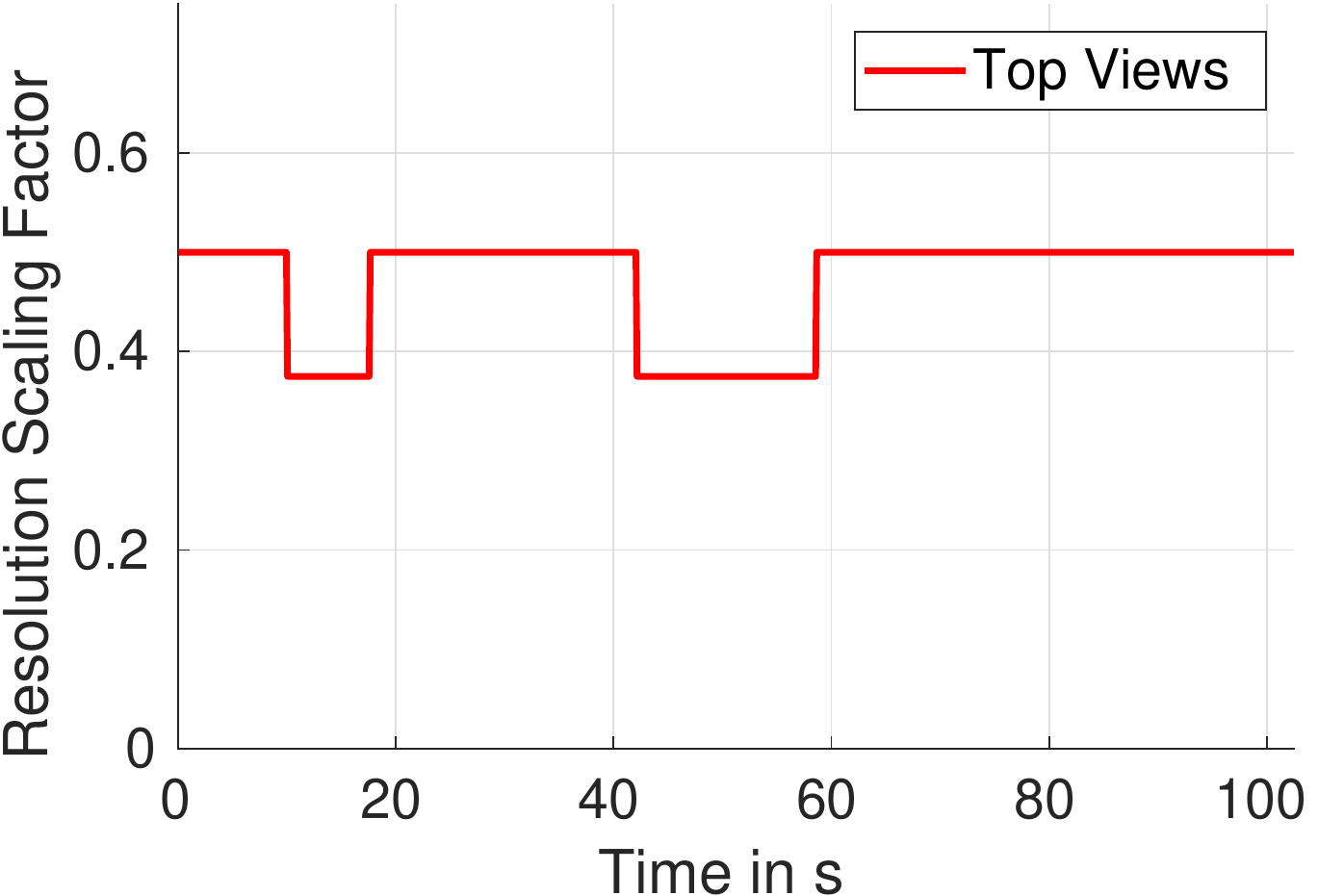}
    }
    \caption{Video Resolution Factors over Time}
    \label{fig:megapixelOverTime}
\end{figure}

%
After the generation of the rate-quality models for all cameras on the vehicle, the framework is put into operation. In the following, insights into the function of mode~A are given for a driving test of approximately~\SI{100}{\second}. The predictions of the available bitrate are artificial, being read from a pre-recorded bandwidth map. As stated in Sec.~\ref{ch:pQoS}, the development and deployment of more sophisticated regression and prediction models is beyond the scope of this paper. \\
Fig.~\ref{fig:bitrateOnGps} shows the total bitrate of all cameras, plotted on the GPS location of the vehicle during the driving test. Great variance of the bitrate can be observed, ranging from~\SI{3000}{} to~\SI{8000}{\kilo\bit\per\second}. The actual camera bitrates over time are plotted in Fig.~\ref{fig:bitrateOverTime}. The front-mounted cameras used for the left and right views have greater~$b_{\text{full}},$ and are always allocated higher bitrates. This is the consequence of the lower visual quality levels achieved for these cameras, due to their orientation, as described in the previous section. The bitrates of the top view cameras are controlled equally at overall lower bitrate levels, compared to the front- and rear-mounted cameras. Given the variation in allocated bitrates, the framework also varies the resolution of the videos. The resolution scaling factors are plotted over time in~Fig.~\ref{fig:megapixelOverTime}. In two occasions, around~$15$ and~\SI{55}{\second}, only lower bitrates are available for several seconds. This results in the selection of lower resolutions for all cameras, except the front-mounted center camera. Between~$70$ and~\SI{90}{\second}, a lower bitrate is predicted as well. However, with the given rate-quality models, the resolutions of the videos are not changed. 

\section{Discussion and Outlook}
The rate-quality models form a good basis to adequately parametrize the videos of the ToD system. However, the rate-quality models were generated from one driving sequence, only. To account for deviations from the models, an assessment of the actual image complexity at runtime is required. This could be done through a pre-processor or feedback, available from the encoder. In addition, extended parameter models could be generated, in dependence on the longitudinal and lateral motion of the vehicle. Alternatively, the generation of rate-quality models for specific driving scenarios is also conceivable. For instance, unprotected left turns yield a greater importance for the front-mounted camera facing to the left. In parking scenarios, the top view cameras, capturing the close surroundings of the vehicle, should be allocated higher priority. \\
Another aspect that can be addressed in future work is the selection of the visual quality metric for generating the rate-quality models. For instance, the~PSNR, which estimates absolute errors, corresponds to the perceived quality to some extent, only. The~MSSIM, which was used in this paper, aims improve this by capturing the similarity of images regarding structural information. In future work, the open question of which metric is the right choice to improve on for the task of ToD will be addressed. 

\section{Conclusion}
This paper presented a flexible video streaming framework for teleoperated driving. It offers the ability to dynamically configure individual video streams of the system. Given the total available uplink bitrate, the framework is capable of automatically handling bitrate allocation and optimization of the video resolution scaling factors. This is based on rate-quality models that were generated for each camera of the system. The operation of the framework is demonstrated on an experimental vehicle with eight cameras. A discussion on the proposed methodology points out several directions for future work and enhancements of the framework. Ultimately, this research aims to explore these possibilities with the goal to further improve the video streaming quality for the application of teleoperated driving.

\section{Appendix}

\begin{figure}[!h]
    \subfigure[Top View Front] {
        \includegraphics[width=0.465\linewidth]{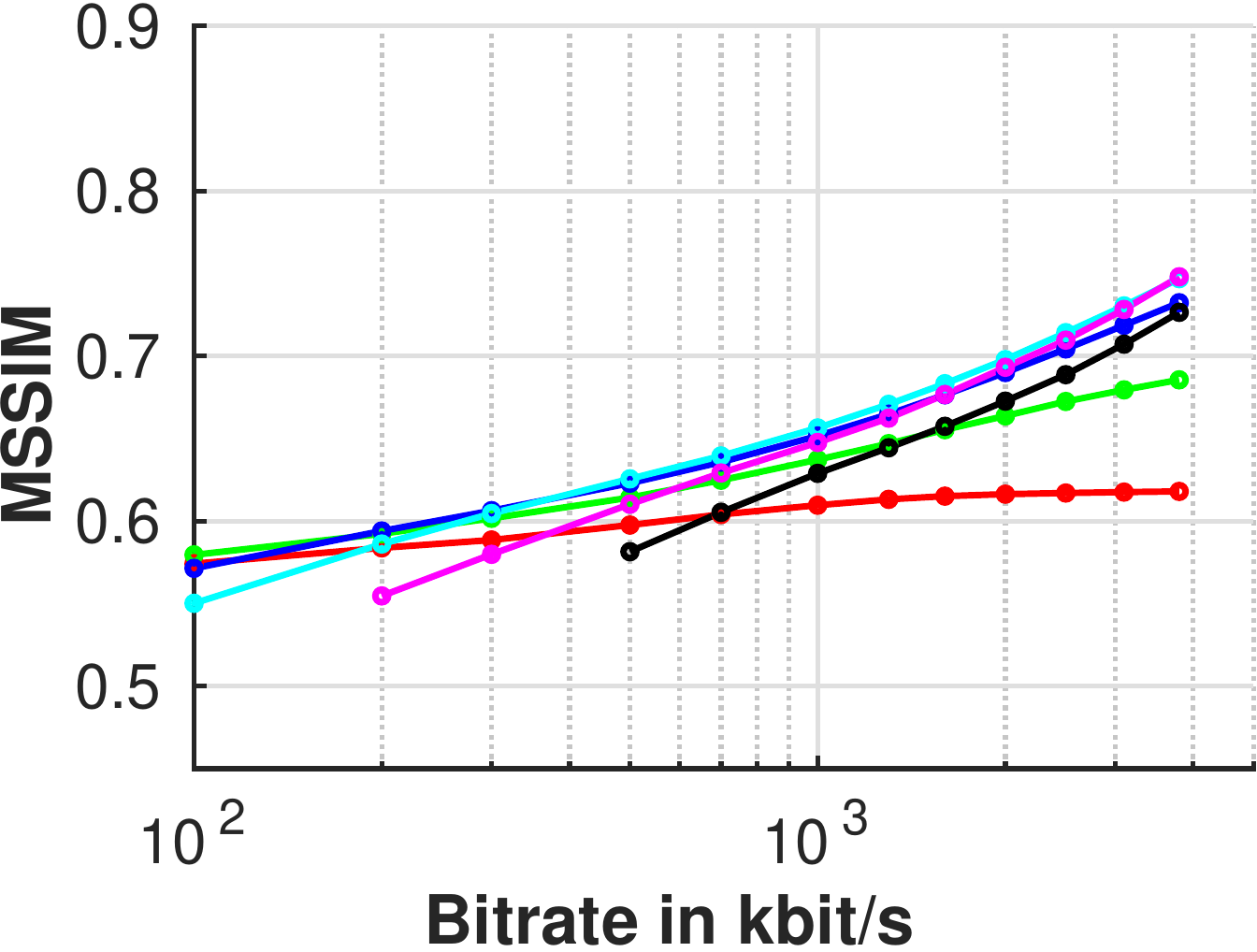}
    }
    \subfigure[Top View Left and Right Average] {
        \includegraphics[width=0.465\linewidth]{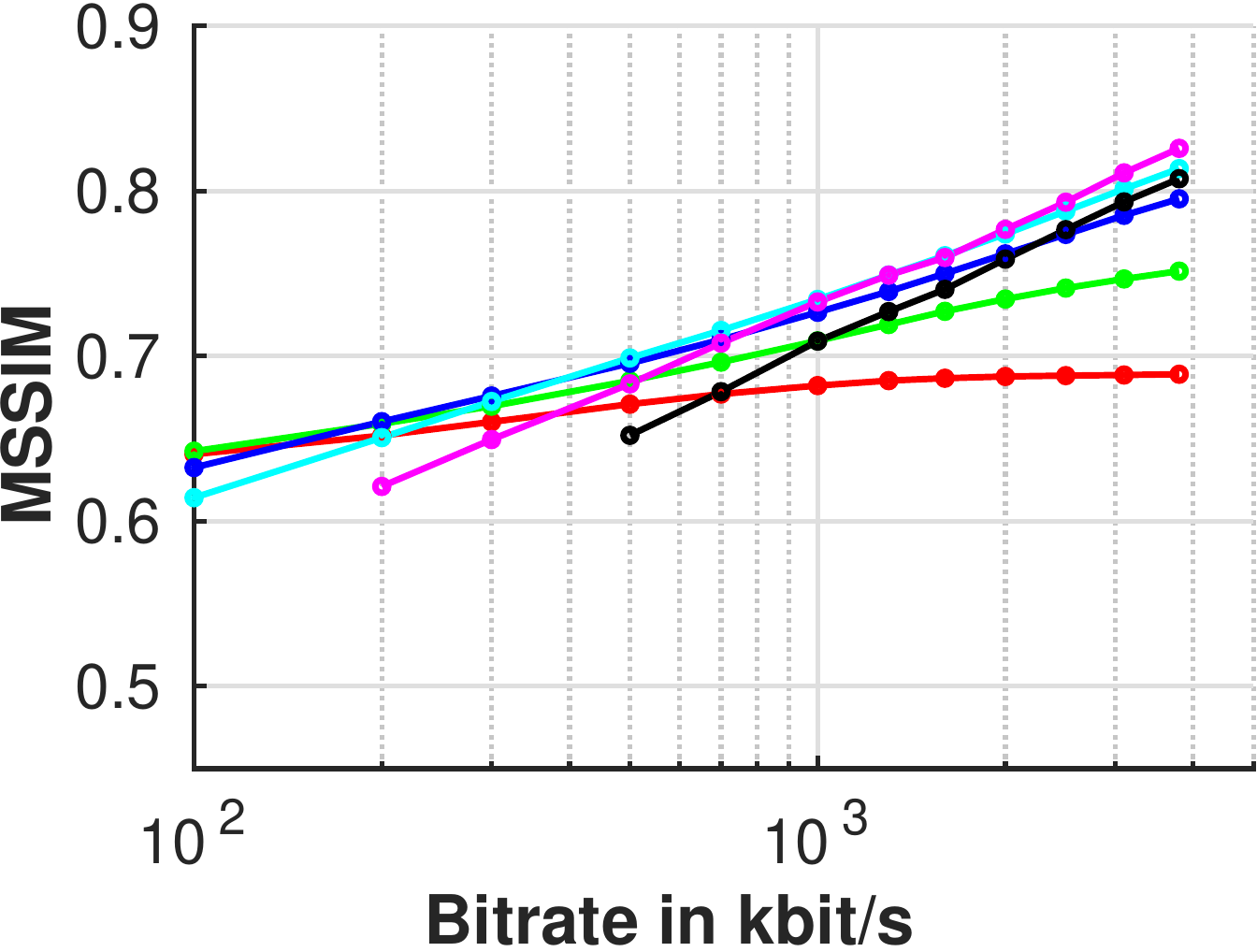}
    }
    \subfigure[Top View Rear] {
        \includegraphics[width=0.465\linewidth]{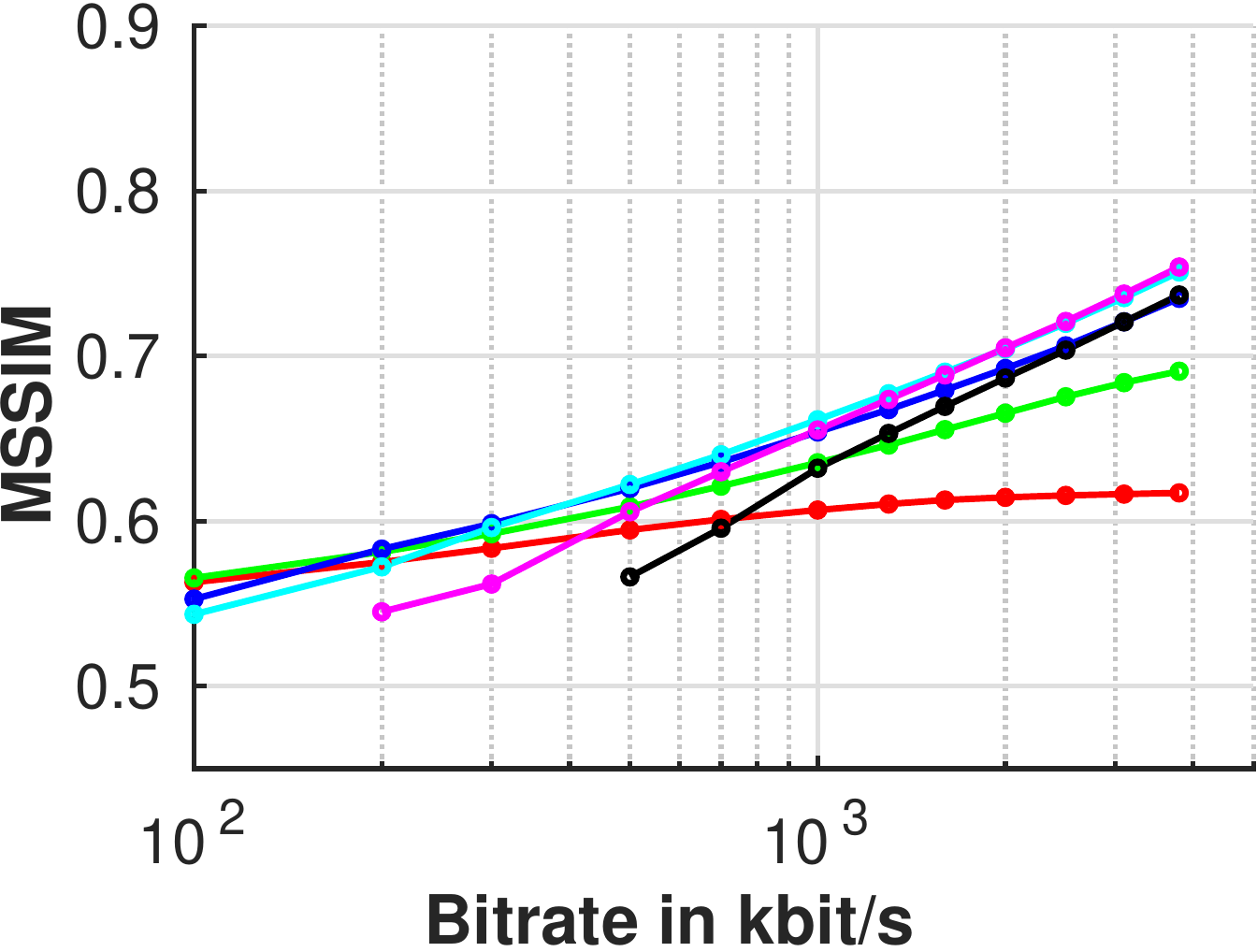}
    }
    \subfigure[Rear-Mounted Center] {
        \includegraphics[width=0.465\linewidth]{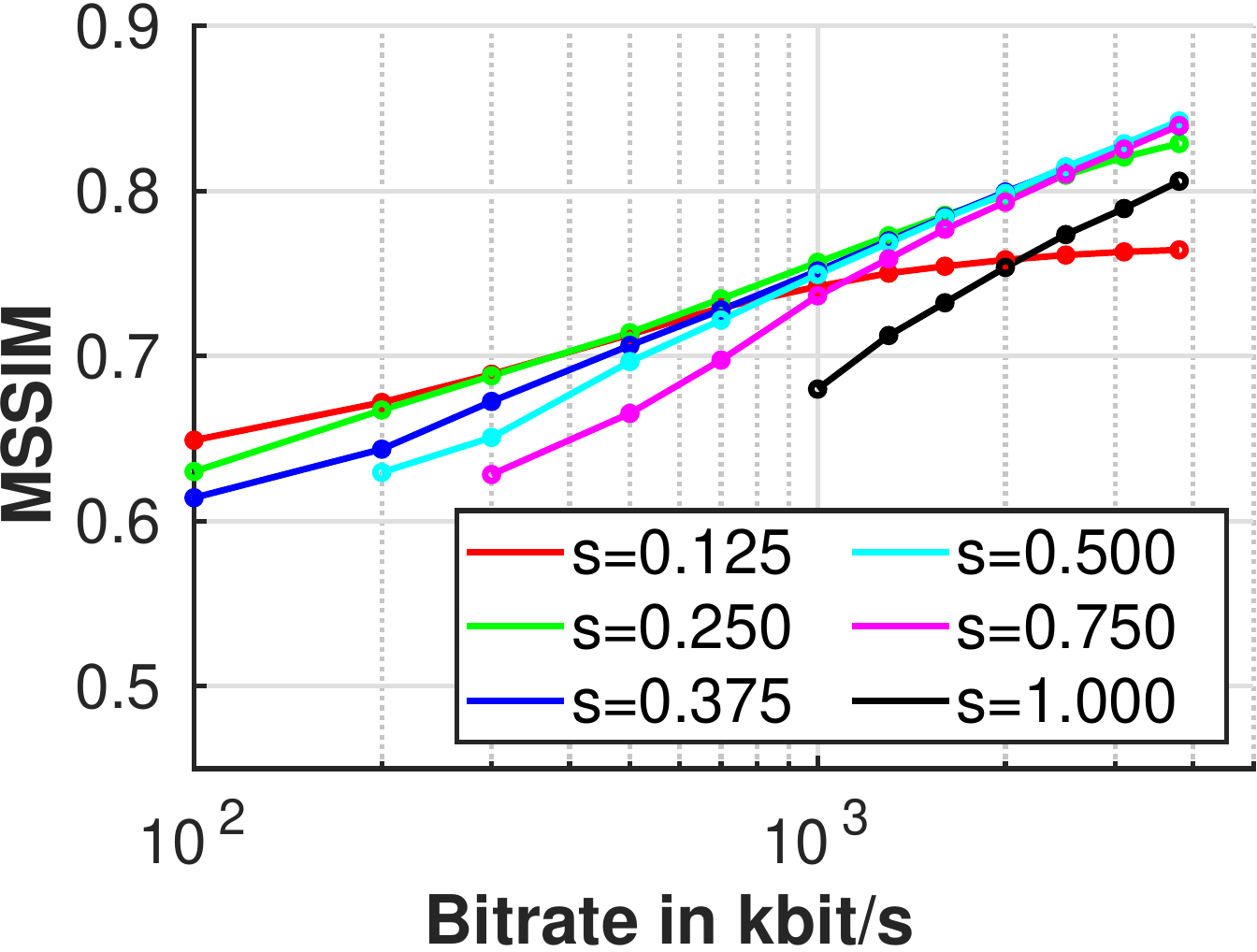}
    }    
	\caption{Rate-Quality Models for other Cameras}
	\label{fig:otherQualCurves}
\end{figure}

\section{Acknowledgements}
This work was presented at the workshop for Road Vehicle Teleoperation~(WS09), IV2021. Andreas Schimpe, as the first author,  was the initiator of the research idea. Simon Hoffmann supported the development and deployment of the framework on the experimental vehicle. Frank Diermeyer made essential contributions to the conception of the research project. The research was partially funded by the European Union~(EU) under RIA grant No.~825050. 

\end{document}